\documentclass[manuscript]{aastex}
\usepackage{graphicx}
\usepackage{epstopdf}
\usepackage{color}
\usepackage{physics}
\usepackage{amsmath}

\slugcomment{Accepted in ApJ}

\shorttitle{Magnetic Field Analysis of AR 11158}
\shortauthors{Johan Muhamad et al.}

\begin{document}


\title{A Study of Magnetic Field Characteristics of Flaring Active Region Based on Nonlinear Force-free Field Extrapolation}


\author{Johan Muhamad\altaffilmark{1}, Kanya Kusano, and Satoshi Inoue}
\affil{Institute for Space-Earth Environmental Research (ISEE), Nagoya University \\
Furo-cho, Chikusa-ku, Nagoya 464-8601 Japan}

\and

\author{Yumi Bamba}
\affil{Institute of Space and Astronautical Science(ISAS)/ Japan Aerospace Exploration Agency (JAXA)\\ 3-1-1 Yoshinodai, Chuo-ku, Sagamihara, Kanagawa 252-5210 Japan}


\altaffiltext{1}{Space Science Center, National Institute of Aeronautics and Space (LAPAN) \\
Jl. Djundjunan No. 133, Bandung 40173 Indonesia}


\begin{abstract}
Coronal magnetic fields are responsible for the onset of solar flares and solar eruptions. However, the type of magnetic field parameters that can be used to measure the critical condition for a solar eruption is still unclear. As an effort to understand the possible condition for a solar flare, we have examined the non-dimensional parameter $\kappa$ introduced by \citet{isi17}, which contains information about magnetic twist distribution and magnetic flux in an active region (AR). We introduce a new parameter $\kappa^\ast$, as a proxy for $\kappa$, and we have analyzed the evolution of $\kappa^\ast$ during the flaring period of an AR using the nonlinear force-free field (NLFFF) extrapolated from the photospheric vector magnetic field data. Using data from the Solar Dynamics Observatory (SDO)/Helioseismic and Magnetic Imager (HMI), we have calculated $\kappa^\ast$ for the AR NOAA 11158 during its three-day flaring period. We found that $\kappa^\ast$ increased to a certain level before two large flares and decreased significantly after their onset. The results suggest that $\kappa^\ast$ may be used as an indicator of the necessary condition for the onset of a solar eruption in the AR. Based on this result, we propose a new method to assess the possibility of a large solar eruption from an AR by combining the parameter $\kappa^\ast$ and information about the magnetic energy of the AR.
\end{abstract}


\keywords{Sun: flares, Sun: corona, NLFFF}



\section{Introduction}

Large solar flares are generated from regions in the photosphere near the magnetic polarity inversion lines (PILs), which contain highly sheared magnetic fields, with the horizontal fields nearly parallel to the PIL. Since the magnetic helicity must be well conserved, the shearing motion and the emerging of sheared magnetic field in the photosphere can carry the helicity to the chromosphere and corona to form high-twist field lines or flux rope \citep{hag84,ber98,pri16}. Previous studies showed that the twist as a component of magnetic helicity can be important since a highly twisted magnetic field can be subject to the helical kink instability \citep{hood79,ein83}. An active region (AR) that contains a highly twisted magnetic field tends to produce a flare when the twist reaches a critical instability state. The threshold for the kink instability may vary depending on the configuration of the magnetic structure. Theoretical studies have shown that in the ideal line-tying condition, the critical threshold for the twist of a flux rope is ${\sim}1.25$ turns or ${\sim}2.5\pi$ radians \citep{hood79}. Previous studies, using simulation of the \citet{tit99} equilibrium, found that the twist of a helical kink unstable coronal configuration must exceed a certain threshold \citep{fan04}. \citet{tor04} found this threshold to be ${\sim}3.5\pi$ radians. However, extracting twist information from observations is very challenging. Several authors have tried to determine the twist via various techniques. For example, \citet{rus96} used the ratio of length to width, \citet{lek05,mal11} employed the best-fit of linear force-free fields to the observations, and \citet{lek07} examined moments of a twist parameter. \citet{bob16} used the mean value of force-free  parameter (alpha) as a proxy for the twist in their prediction method, but they found that it was not a significant predictor for coronal mass ejection (CME). \citet{lea03} argued that the pre-flare coronal fields in ARs rarely have sufficient twist for the kink instability.  However, from their study of many sigmoids, \citet{rus05} showed observational evidence for the kink instability in solar filament eruptions.   

Recent studies have attempted to calculate the twist of the field lines from a nonlinear force-free field (NLFFF) model based on the photospheric magnetic field data \citep{ino11,ino13,ino16,guo13,liu16}. With a NLFFF model, one can calculate the twist of each field line to determine the distribution of twist for the whole AR. However, it is still debatable how to define the twist of a flux rope that consists of many field lines \citep{liu16}. \citet{pev14} suggested that a proper definition of magnetic twist for an AR should be normalized by the magnetic flux, since the number of field lines is infinite. \citet{liu16} showed that an NLFFF model can reconstruct a high-twist magnetic flux rope that exceeds the kink instability threshold before a flare. They suggested that the twist of the flux rope axis can be used as a parameter in forecasting solar eruption. Several studies based on the NLFFF models for different ARs have shown that the pre-flare coronal fields are weakly twisted \citep{bob08,ino14b,ino15}. \citet{ino15} pointed out that a triggering mechanism associated with magnetic reconnection may make it possible for a magnetic structure with a twist less than the kink instability threshold to produce a flare. Therefore, it may not be necessary for a magnetic structure in an AR to have a twist greater than 1.25 turns in order to produce a flare, even if the kink instability works as the driver for the main phase of the flare.

Other theories have been developed to explain the mechanism of solar flare. When two magnetic arcades are close to each other, tether-cutting reconnection near the footpoints of the loops can cause an eruption \citep{moor01}. The creation of magnetic flux rope in the corona can happen through this tether-cutting mechanism or flux cancellation in the photosphere \citep{sch15}. According to the magnetic break out model, reconnection can also happen in the null points when magnetic loops interact with an overlying field that will eventually lead to an eruption \citep{ant99,wyp17}. A solar eruption can also be caused by torus instability (TI) when the current inside the flux tube is strong enough to generate hoop force that is larger than the strapping force from the magnetic field surrounding the flux tube \citep{kli06,dem10}. In this kind of instability, how fast the external magnetic field change with height, which is defined by the decay index ($n$), is important to determine the critical condition of the eruption. \citet{dem10} confirmed previous studies by \citet{bat78} and \citet{kli06} that the TI will grow when $n>$1.5. The constraint of the decay index can prevent the eruption (CME) although the magnetic twist of the flux rope exceeds the critical limit for kink instability \citep{ama18}. Recently, \citet{isi17} studied the stability of a double-arc current carrying loop that can be formed by a tether-cutting reconnection between two loops and introduced the double-arc instability (DAI). Since the driving force of the DAI is the hoop force acting on the loop as same as the TI, these two instabilities can be categorized into the same group, which can be called the ``hoop-force-driven" instability. However, the magnetic structure for the DAI is more complex than a simple torus. Therefore, the critical parameter of the DAI is not same as that of the TI. The critical condition of the DAI is given by a new parameter $\kappa$, whereas the critical condition of the TI is determined by the decay index $n$. Because of that, the magnetic structure is able to be unstable to the DAI even though the external field is not decayed and the DAI can grow in a portion lower than the critical height of the TI as shown by \citet{isi17}. On the other hand, although the DAI can drive the initial phase of flare, the TI may play a crucial role to make a full eruption and CME, because the DAI can explain only the process in which the double-arc loop formed by the tether-cutting reconnection expands to a simple torus. 

Several studies of flare trigger mechanisms have shown that both highly sheared field and reconnection among some field lines are required in order to produce a flare \citep{kus12,muh17}. These studies used MHD simulations to show that both the configuration of the pre-existing magnetic field near the PIL and the orientation of small-scale magnetic disturbances on the PIL are important for triggering a flare. Therefore, having a large free energy accumulated in a strongly sheared magnetic field is necessary, but not sufficient for producing a flare.  This result is consistent with the theory of the DAI. From numerical analyses of the stability of a double-arc electric current loop, \citet{isi17} suggested that the magnetic twist and the reconnected flux may play complementary roles in triggering a flare. If a double-arc loop can be formed by a reconnection between two sheared field lines, as in the tether-cutting scenario \citep{moor01}, the $\kappa$ parameter defined as
\begin{equation}
\kappa = T_w \frac{\phi_{rec}}{\phi_{tot}} ,
\end{equation}
can be used to measure the critical condition of the DAI, where $T_w$ is the twist of a magnetic loop and $\phi_{rec}$/$\phi_{tot}$ is the fraction of the reconnected flux over the total flux crossing over the considered PIL.

In their study, \citet{isi17} found that when the value of $\kappa$ exceeds a threshold $\kappa_0$, a double-arc electric current loop becomes unstable. The threshold $\kappa_0$ varies from approximately 0.08 to 0.175, depending on the external magnetic field enveloping the double-arc structure. This suggests that the magnetic twist and tether-cutting reconnection are important in destabilizing sheared magnetic field \citep{isi17}. However, calculating the value of $\kappa$ for an AR is not trivial, because the observational data lack information about the magnetic twist and the reconnected flux. Although one can determine the force-free alpha parameter from magnetogram data, those data contain no information about the length of magnetic loops that can be used to derive the twist. Moreover, the reconnected flux, which is required to calculate $\kappa$, is hardly known from observations. Therefore, it is important to develop a way to calculate $\kappa$ in an AR to understand the characteristics pattern of its evolution before and after a large flare happens.   

We have employed DAI analysis to study flares in the AR NOAA 11158. The evolution of the magnetic field in this AR have been extensively studied, e.g. regarding the magnetic structure \citep{sun12,tor13,zhao14,mal14}, magnetic helicity \citep{jin12,dal13,tzi13,zha16}, photospheric field \citep{liu12,pet12,wan12,pet13}, magnetic energy \citep{sun12,asc14,mal14}, magnetic twist \citep{liu13,sun12,ino11,ino14b,zhao14,mal14}, and others. Observational studies suggested that tether-cutting reconnection happened in the AR during the M6.6 \citep{liu12} and X2.2 \citep{wan12,liu13} flares. This suggests that double-arc current structure occurred in the AR, and thus DAI analysis can be applied to explain the occurrences of these flares. Our goal is to find a parameter that can be used to study the capability to produce a solar flare by implementing such a DAI analysis for a real AR. First, we study the spatial and temporal evolution of the twist distribution of the AR based on the NLFFF model. Next, we calculate a parameter to enable us to extract information about the twist and to determine the reconnected flux. Finally, we show how our proposed parameter can be used as a possible forewarning for the likelihood of flare occurrence in an AR. This paper is organized as follows: Data for the AR and the method we used in our study, are described in Section 2. Section 3 presents the results of the NLFFF extrapolations and describes the evolution of the twist and other parameters. There, we also use our results to discuss some issues related to the necessary condition of the AR for a flare. Finally, we summarize and conclude our paper in Section 4.

\section{DATA AND METHOD} 
\subsection{Observed Data and the NLFFF Model}

We have used Solar Dynamics Observatory (SDO)/Helioseismic and Magnetic Imager (HMI) observations of the vector magnetic field from the Spaceweather HMI Active Region Patch (SHARP) \citep{bob14}. SHARP data have a 12-minute cadence of observations, in which the magnetic flux has been remapped to a Lambert Cylindrical Equal-Area (CEA) projection\footnote{The HMI SHARP (CEA) data (hmi.sharp$\_$cea$\_$720s) is available on the Joint Science Operation Center (JSOC), Stanford University (http://jsoc.stanford.edu/).}. These data contain the three components of the vector magnetic field ($B_r$, $B_\phi$, $B_\theta$) obtained from a very fast inversion of the Stokes vector using Milne-Eddington model for the solar atmosphere \citep{bor11}. The 180-deg ambiguity is resolved by using the minimum-energy method \citep{lek09}. We used a three-hour cadence of the SHARP data for AR 11158, extending from 00:00 UT on 2011-Feb-13 to 21:00 UT on 2011-Feb-15 with 744$\times$377 pixels covering about 268$\times$134 Mm$^2$. For detailed analyses of two large flares that occurred during this period, we used the 12-minute cadence of the SHARP data from 15:00$-$18:00 UT 2011-Feb-13, when M6.6 flare happened, and 00:00$-$03:00 UT 2011-Feb-15, when X2.2 flare occurred. 

We reduced the original vector magnetogram data by using IDL CONGRID function with the nearest-neighbour interpolation to 256$\times$128 pixels, which we used as bottom boundaries for extrapolating the coronal magnetic field. We employed the NLFFF extrapolation technique to  reconstruct the coronal magnetic field by applying the magnetohydrodynamics (MHD) relaxation method \citep{ino14a} in a computational box of 256$\times$128$\times$128 grids. First, we calculated the potential field for the AR from the normal component ($B_r$) of the given boundary condition using the Fourier method \citep{ali81}. The tangential components ($B_\phi$, $B_\theta$) then incrementally changed at the bottom boundary to match the observational tangential components. Next, we calculate a set of MHD-like equations until the solution approaches the force-free field (for the details of MHD relaxation method see \citet{ino14a}). All the physical variables at all the boundaries are fixed during the iteration. The method and parameters used here are similar to the NLFFF calculations in our previous work \citep{muh17}. 

\subsection{Calculating $\kappa^\ast$-Parameter}
To derive information similar to that embodied in the parameter $\kappa$ from the observational data, we propose the alternative parameter namely $\kappa^\ast$ that is defined as follows:
\begin{equation}
\kappa_{T_c}^\ast = \frac{\int_{_{\frac{T_w}{T_c}>1}} \abs{T_w}\ \mathrm{d}\phi}{\phi_{tot}} ,
\end{equation}
with $\mathrm{d}\phi=\abs{B_r}\mathrm{d}S$.

In this formula, instead of using the reconnected flux, which is not measurable, we integrate the magnetic twist only over the high-twist area as a function of magnetic flux ($\phi$). The threshold for the integrated magnetic twist is given by $T_c$. The concept of replacing the reconnected flux with the high-twist flux is based on the assumption that internal reconnection between high-twist field lines in the pre-flare phase is more effective in enhancing the creation of higher twist flux ropes, which can be expected to drive a flare \citep{dem10}. This assumption is supported by the fact that flares usually come from the core of an AR, which contains highly twisted field lines, and which stores a large free energy. In fact, the NLFFFs of several ARs showed that the shapes and locations of flare ribbons are very well matched with the footpoint locations of the highly twisted field lines \citep{ino13,ino16,liu16}. Therefore, it is likely that a flare is triggered by the reconnection of high-twist field.
 
The definition of $\kappa_{T_c}^\ast$ requires a proper threshold ($T_c$) to determine the high-twist area, since this can affect the overall estimation of $\kappa_{T_c}^\ast$. \citet{ino14b} found that the maximum twist value of the AR 11158 was less than unity before the flare. A few of field lines had twist more than 0.5, although most of the field lines had smaller values. \citet{ino11,ino16} showed that ARs 10930 and 12192 also had similar magnetic twist properties. Based on those results (\citet{ino11,ino14b,ino16}) we have employed 0.5 as a benchmark for the high-twist threshold and have surveyed the dependence of $\kappa^\ast_{T_c}$ on $T_c$ around this value (0.5). Another possible way to determine $T_c$ is by analyzing the twist distribution after a flare occurs. This distribution can be expected to change, because the flare relaxes the magnetic twist. Magnetic field will thus tend to have lower twist values after a flare. Magnetic flux that loses twist can be considered to be reconnected flux. Therefore, the maximum twist value of the field remaining after a flare, denoted by $T_m$, can be used as a minimum threshold for determining the high-twist field before the flare. Here, we first try to calculate $\kappa_{0.5}^\ast$, and we then compare it with $\kappa_{T_m}^\ast$, which we compute using coronal magnetic field reconstructed from NLFFF extrapolation. The twist of a field line is calculated from the following definition of magnetic twist \citep{ber06}:
\begin{equation}
T_w = \frac{1}{4\pi} \int \alpha\ \mathrm{d}l .
\end{equation}
Here, alpha is the force-free parameter 
\begin{equation}
\alpha = \frac{\nabla \times \mathbf{B} \cdot \mathbf{B}}{|B|^2} ,
\end{equation}
which is calculated for each point in a computational box along a field line, where $l$ is the length coordinate along the field line. The magnetic twist is thus defined as a function of each field line. This definition was also used by \citet{ino15,ino16} to calculate the twist. We then evaluate equation (2) for several different values of $T_c$ at each different epoch.  

In the weak field region, there are big numerical errors in the calculation of $\alpha$, and the evaluation of the magnetic twist is less reliable. Therefore, we exclude the magnetic field lines coming from weak field regions by using a minimum field threshold in the calculation. In this work, we only consider magnetic flux density larger than 140 G in calculating $\kappa_{T_c}^\ast$. The total flux ($\phi_{tot}$) in equation (2) is defined as the total unsigned flux, which is the closed field within the core of the AR. This excludes fluxes for which the field lines are open or their connectivity is beyond the area within the core of the AR, where the center of the flare ribbons was observed during the flares.

\section{RESULTS AND DISCUSSION}
\subsection{NLFF Results}

\begin{figure}
\epsscale{1.0}
\plotone{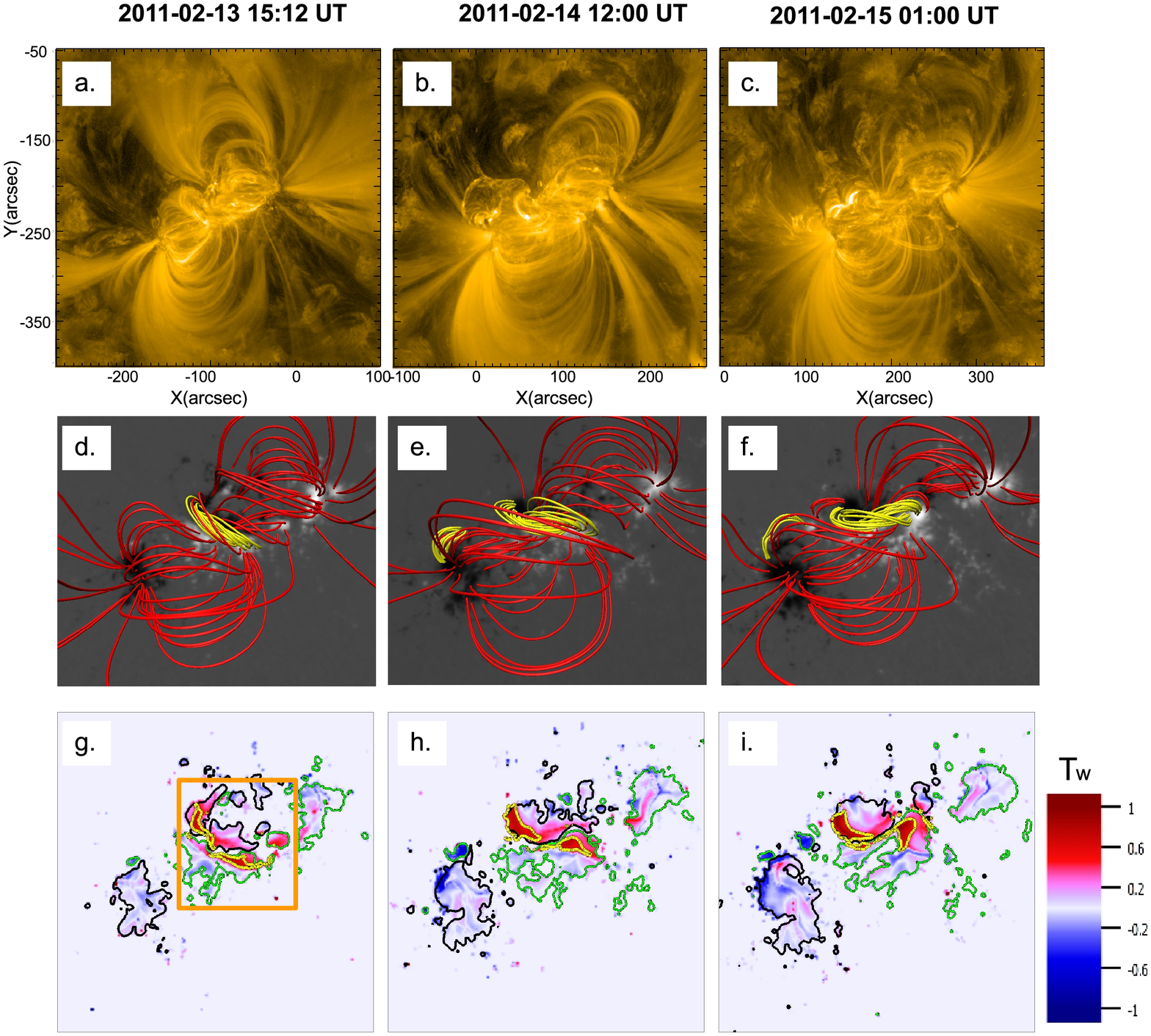}
\caption{The coronal magnetic fields of AR 11158 observed by SDO/AIA 171 ${\AA}$ at three different times (a),(b), and (c). The coronal magnetic fields reconstructed from NLFFF extrapolations for AR 11158 corresponding to the epochs of the top panels are presented in (d),(e), and (f).  The yellow (red) lines show selected magnetic field lines with twists higher (lower) than 0.5 turn. Twist map for AR 11158 calculated from the NLFFF model (g), (h), and (i). The green (black) contours in the bottom panels represent the normal components of magnetic field at 560 G (-560 G), respectively. The yellow contours show the integration surface for calculating $\kappa_{T_c}^\ast$ with $T_c = 0.5$.  
\label{fig:f1}}
\end{figure}

AR 11158 initially occurred as two simple bipoles before growing to become complex quadrupole systems. Rotational and translational motions of the spots were observed within the AR \citep{jia12}; these caused the magnetic field near the PIL to become sheared and elongated \citep{tor14}. Figures \ref{fig:f1}(a), (b), and (c) show the coronal magnetic field of AR 11158 observed by AIA 171 ${\AA}$ at three different times. AR 11158 produced many flares and CMEs during its passage across the solar disk. In this paper, we focus on the M6.6 flare, which occurred on 2011 February 13 at 17:28 UT, and on the X2.2 flare, which happened on 2011 February 15 at 01:44 UT. These flares occurred in the core of the AR, which is marked by the orange rectangle in Figure \ref{fig:f1}(g).   
  
NLFFF extrapolation provides the reconstructed coronal magnetic field for AR 11158 from 2011 February 13 $-$ 2011 February 15 (Figures 1(d), (e), and (f)). The field lines (yellow lines) within the core of the AR have arcade-like structure with a relatively strong twist mainly near the PIL. These figures show that the magnetic field evolved during this period. Although we did not quantitatively compare the field lines with the observation, in general, the reconstructed coronal fields morphologies match with the observations in Figures 1(a), (b) and (c). The general morphologies and the locations of the high-twist fields also in agreement with many previous studies \citep{jin12,sun12,dal13,liu13,wan13,ino13,ino14b,asc14,zhao14,mal14}. Unlike \citet{zhao14} who could identify the twisted flux rope from the topology of the reconstructed coronal field, we could not find obvious topological signature of flux rope existing in our NLFFF during our analysis time window. It might be due to the fact that there were few magnetic flux with twist higher than one-turn in our NLFFF and it is difficult to be topologically defined as a flux rope. However, our result is consistent with other NLFFF results \citep{jin12,sun12,liu13,wan13,ino14b,mal14}. The high twist region in our result also in agreement with the region with high helicity flux \citep{dal13} and the location of the flare ribbons \citep{bam13,liu13}, as well as the high current density region \citep{jan14}. Figures \ref{fig:f1} (g), (h), and (i) show the evolution of the twist distribution map, with the magnetic twist of the field lines plotted at the footpoints of field lines according to a color scale. This shows that the high-twist (strongly right-handed twist corresponding to $T_w>0.5$) areas are concentrated in only a limited part of the AR. The high-twist area grew and became even more twisted just before the X2.2 flare (Figure \ref{fig:f1}(i)). Most parts of the AR have twist values less than 0.25, but near the PIL the twist can reach more than 0.5, even up to about a full turn. This is consistent with the results of \citet{sun12} and \citet{ino14b}. A high-twist (strong negative/left-handed twist) area also developed in the eastern part of the AR, which did not exist initially on February 13. Both of these high-twist areas produced several flares. However, here we focus on the flares resulted from the high-twist core region near the center of the AR, where the M6.6 and X2.2 flares occurred.

\subsection{Twist Distribution and Evolution}
\subsubsection{2011 February 13 (M6.6 Flare Event)}

In order to examine the evolution of twist before and after M6.6 flare, we plotted the distribution of magnetic twist in the core of AR 11158 from 15:00$-$18:00 UT on 2011 Feb 13 with a 12-minute cadence. Note that M6.6 flare started at around 17:28 UT. We obtained the twist distribution by binning the absolute value of twist with increments of 0.05 and then summing the magnetic flux of the $B_r$ component, which corresponds to the footpoints of the field lines for which we calculated the twist within the area in Figure \ref{fig:f1}. We excluded field from areas with weak magnetic field (less than 280 G) in order to eliminate noise. The plot of the amount of magnetic flux as a function of its corresponding twist, normalized by $7.72 \times 10^{21}$ Mx, is presented in each panel of Figure \ref{fig:f2}, which shows the evolution of the twist distribution of AR 11158 before and after M6.6 flare. In general, most field lines had only small amount of twist, with values less than 0.2 turn. However, several field lines had a twist exceeding 0.5 or even 0.8 turn before the flare occurred. After the flare, almost all twists of the high-twist magnetic field gradually reduced to less than 0.6 turn. This result is consistent with the result of \citet{liu12}. 

The selected spatial evolution of the magnetic twist distribution in AR 11158 during the three-hour period from 15:00$-$18:00 UT on February 13 is presented in Figure \ref{fig:f3}. This figure shows how the high-twist (larger than 0.5 turn) field area (enclosed by the green contour) intensified, with the magnetic flux was $1.6 \times 10^{21}$ Mx, just before the flare (17:12 UT) and then shrank significantly to $3.6 \times 10^{20}$ Mx after the flare (18:00 UT). This suggested that magnetic helicity might have been transferred from the core of the AR to the upper region where the magnetic field erupted during the flare to the interplanetary space. The results of the twist distribution and evolution before and after the flare are consistent with the result of \citet{ino13}. 

\subsubsection{2011 February 15 (X2.2 Flare Event)}
X2.2 flare started around 01:44 UT on February 15. To study the evolution of magnetic twist in AR 11158 before and after the flare, we plotted the twist distribution in the AR from 00:00$-$03:00 UT on February 15 with a 12-minute cadence. Selected plots of the twist distribution are shown in Figure \ref{fig:f4}. Over this time period, the trend was similar to the development of the twist distribution for the M6.6 flare. However, before the X2.2 flare, more field existed with twist higher than 0.5, or even larger than 0.8 turn. This shows that, during this period, the magnetic field of AR 11158 had stronger shear than before the previous event. Field with twist higher than 0.8 turn mostly disappeared after the flare. This is also similar to the previous event, although the AR retained more flux with magnetic twist higher than 0.5 even one hour after the peak of the flare.    

We also plotted contours for magnetic field with twist higher than 0.5 over the $B_r$ component of the magnetogram data for the period 00:00$-$03:00 UT on February 15 with a 12-minute cadence. Selected plots of these images are shown in Figure \ref{fig:f5}. The region with magnetic twist higher than 0.5 did not exhibit significant changes before the flare ($2 \times 10^{21}$ Mx at 01:36 UT), but it gradually reduced in size just after the flare ($1.9 \times 10^{21}$ Mx at 02:12 UT). However, high-twist region with $T_w > 0.6$ decreased from $1.45 \times 10^{21}$ Mx before the flare (01:36 UT) to $7 \times 10^{20}$ Mx after the flare (02:12 UT). Unlike M6.6 flare, high-twist flux still remained in relatively large area, instead of significantly disappearing. This implies that AR 11158 still retained a significant amount of free energy. These results of the twist distribution and evolution before and after the flare are also consistent with the result of \citet{ino13}.

\subsection{Evolution of $\kappa^\ast$}

After we calculated and analyzed the twist, we calculated $\kappa_{T_c}^\ast$ using equation (2) with different thresholds ($T_c$). We used the first threshold, 0.5, as a benchmark to explore the general trend of the evolution. The evolution of the GOES X-ray flux, $\kappa_{0.5}^\ast$, and the total magnetic energy are shown in Figures \ref{fig:f6} (a), (b), and (c), respectively. The three-hour cadence data are used to make the evolutionary plot of $\kappa_{0.5}^\ast$ and the total magnetic energy. We used the 12-minute cadence data around the time when the two big flares occurred in the core of the AR, on February 13, from 15:00$-$18:00 UT and on February 15, from 00:00$-$03:00 UT. Note that $\kappa_{0.5}^\ast$ increased before the flares and  decreased abruptly just after the flares.

For a more detailed analysis, we have plotted in Figures \ref{fig:f7} (a) and (b) the evolution of the GOES X-ray flux during M6.6 and X2.2 flare, respectively. We also plotted in Figures \ref{fig:f7} (c) and (e) the evolution of $\kappa_{T_c}^\ast$ and free energy, each with a 12-minute cadence, just before and just after the M6.6 flare, respectively. We used three different values for the threshold ($T_c$) for the high-twist flux for this event (shown as different colored lines in Figure \ref{fig:f7} (c)). The first threshold ($T_c=0.5$; red line) is the benchmark. We obtained the second threshold ($T_c=0.6$; green line) from the highest value of the magnetic twist after the flare ($T_m$), as can be seen from the last panel in Figure \ref{fig:f2}. The third threshold ($T_c=0.55$; blue line) is just the mean value between the first and second, which we used for comparison. Figure \ref{fig:f7}(c) shows that $\kappa_{T_c}^\ast$ increased for each value of $T_c$ before the flare and then significantly decreased just after the flare started. The trend is similar for all three thresholds, although $\kappa^*_{T_c}$ becomes smaller for higher $T_c$. It is interesting that the maximum $\kappa_{T_m}^\ast$ is slightly greater than 0.1. This is consistent with the theoretical threshold for $\kappa$ in DAI analysis done by \citet{isi17}. Note that the free magnetic energy decreased slightly after the flare, which shows that some energy was released during the flare.    
 
In Figures \ref{fig:f7}(d) and (f), we plotted the same parameters as in Figures \ref{fig:f7}(c) and (e), but for X2.2 flare. We also used three different thresholds ($T_c$) for the high-twist flux (shown as different lines in Figure \ref{fig:f7} (d). The first threshold ($T_c=0.5$; red line) is again the benchmark threshold. The second ($T_c=0.75$; green line) is the highest twist of the field after the flare ($T_m$) (see the last panel in Figure \ref{fig:f4}). The third threshold ($T_c=0.625$; blue line) is the mean between the first and second, which we have again used for comparison. Note that $\kappa_{T_c}^\ast$ increased long before the X2.2 flare occurred. Figure 6(b) shows that this increase began on February 14 around 00:00 UT and exceeded 0.2 around 04:00 UT, which corresponded to the phase when the AR resumed the flare activity producing the C8.3 flare at 04:29 UT on February 14. However, after the X2.2 flare happened, $\kappa_{T_c}^\ast$ decreased significantly, just as it did following the M6.6 flare event. Note also in Figure \ref{fig:f7}(d) that $\kappa_{T_m}^\ast$ (green line) has a maximum value slightly below 0.1. This value is again consistent with the critical threshold for the DAI introduced by \citet{isi17}. Figure \ref{fig:f7}(f) shows that the free energy was at the higher level before the X2.2 flare than before the M6.6 flare. It increased during the flare, and finally decreased after the X2.2 flare. 

\subsection{Determination of $T_c$}
We have found that the twist of magnetic field in the AR evolved significantly during our three-day analysis time window. In general, the magnetic field of the AR tended to relax to a lower state after the flares occurred. Because the coronal plasma just above the core of the AR erupted, it carried some of the high-twist field from the lower corona to the higher corona or out into interplanetary space. Subsequently, the high-twist magnetic field in the core of the AR relaxed and the magnetic twists decreased. However, some of the high-twist magnetic field still remained there even after the flares. This result that magnetic twist decreased after the flares is consistent with previous studies by \citet{ino13} and \citet{liu16}. 

Previous studies of flare ribbons using footpoint displacement analyses in MHD simulations \citep{tor13,ino15,muh17} and quasi separatrix layers (QSL) in NLFFF models \citep{aul12,guo13,jan14,sav16} have shown that the shapes and locations of flare ribbons usually correspond to the shapes and locations of the footpoints of the reconnected flux. Here, we have found that the contours and locations of the high-twist fields in Figures \ref{fig:f3} and \ref{fig:f5} correspond to the shapes and locations of the observed flare ribbons  for both flares in Figures \ref{fig:f8} (b) and (d). During the flares, reconnection thus tends to involve high-twist fields. This result is consistent with the study by \citet{ino15}. It also supports the idea that the reconnected flux in equation (1) can be replaced by some part of the high-twist flux in equation (2), although they are not completely identical. Moreover, the flare ribbon marks the reconnected flux during the flare, while the reconnected flux in equation (1) only includes internal reconnection among the field lines before the flare. Therefore, the high-twist flux always overestimates the reconnected flux in the pre-flare state. Consequently, $\kappa_{T_c}^\ast$ generally has a higher value than the initial definition of $\kappa$.

The evolution of $\kappa_{T_c}^\ast$ in Figures \ref{fig:f7}(c) and (d) suggests that the AR tended to have relatively large value of $\kappa_{T_c}^\ast$ before a flare and it quickly decreased after the flare. These results suggest that we may be able to use $\kappa_{T_c}^\ast$ as a useful parameter as a forewarning of a flare. However, it is still not clear how we can choose the threshold ($T_c$) for the magnetic twist to use it as a flare predictor. For both of the flares we studied, when $T_c = T_m$, the values of $\kappa_{T_c}^\ast$ just prior to the eruption were around 0.1, which is consistent with the critical value derived by \citet{isi17}. They found that the critical $\kappa$ for the DAI were 0.08 and 0.125, respectively, for a point source and for the external field that decay exponentially with altitude. We suggest that the maximum value of the remaining high-twist field ($T_m$), has more physical meaning than other values of twist as a proper value of $T_c$ for the high-twist flux. Using this threshold reduces the possibility of including non-reconnected flux in calculating $\kappa_{T_c}^\ast$, which makes it more consistent with the theoretical definition of $\kappa$.

\subsection{Importance of Trigger Structure}
Our results show that $\kappa_{T_c}^\ast$ increased about one day before the X2.2 flare, while it increased only about 6 hours before the M6.6 flare. In order to understand the different features of these two flares, we have investigated their trigger mechanisms by analyzing the magnetogram data taken by Hinode/Solar Optical Telescope (SOT). Figure \ref{fig:f8} shows the evolution of the core of the AR before and after the M6.6 and X2.2 flares. 

\citet{bam13,bam14} found that M6.6 flare was triggered by a relatively large magnetic ``peninsula", in which a region with positive polarity intruded into the one with negative polarity (the yellow circle in Figure \ref{fig:f8}(a)). The trigger structure (peninsula) was formed since around Feb 13 13:00 UT by inflow of small positive magnetic patches from the northern positive region \citep{tor13}. This caused the azimuth angle of the magnetic field on the west (the right side) boundary of the peninsula to be reversed from that of the main sheared field in the PIL, enabling reconnection on that boundary to occur by means of the tether-cutting mechanism. This kind of trigger structure has been defined as a ``reversed-shear" (RS) structure by \citet{kus12}. This suggests that in the M6.6 flare, the magnetic structure of the core of the AR was strongly perturbed by the presence of the large peninsula structure before the flare just after the $\kappa_{T_c}^\ast$ increased significantly. Such a large peninsula structure can easily trigger a flare if the magnetic twist grows high enough, and this provides a possible explanation for the prompt onset of M6.6 flare immediately after the value of $\kappa_{T_c}^\ast$ exceeded the threshold. 

On the other hand, X2.2 flare was triggered by a small RS-structure near the PIL (the yellow circle in Figure \ref{fig:f8}(c)). This small structure could hardly trigger a flare, since it could only perturb a small portion of the lower coronal field \citep{bam13}. In addition, we found that the formation of the trigger structure (peninsula) for the X2.2 flare just started from around Feb 14 21:00 UT, which was about one day after the $\kappa_{T_c}^\ast$ increased significantly. Hence, although helicity was continuously injected into the core of the AR, the field structure was able to maintain its high $\kappa_{T_c}^\ast$ value for a relatively long period before the structure of the small RS field met the condition for instability. For that reason, more free energy and magnetic helicity were stored before the onset of the flare, and finally a bigger flare erupted than the previous one. This suggests that the structure of the trigger field is also important to complement the information provided by $\kappa_{T_c}^\ast$ for estimating the probability of the occurrence of a flare.   

\subsection{Magnetic Energy}
        
Analysis of the magnetic energy of the core region shows that the total energy increased continuously during the three-day period covered by the data we used, as can be seen in Figure \ref{fig:f6} (c).  Although from Figure \ref{fig:f6}(c) we see that the magnetic energy accumulating in the AR increased continously during the three days of observation, the $\kappa_{0.5}^\ast$ profile shows dramatic changes before and after the flares. This suggests that $\kappa_{0.5}^\ast$ can serve as a proxy for the degree of DAI necessary for an AR to produce a flare. On the other hand, the total magnetic energy did not show significant changes even during the flares, and instead increased almost monotonically. Ultimately, substantial amount of energy about $1.2\times10^{32}$ erg was stored before the X2.2 flare after the M6.6 flare. This implies that the amount of magnetic energy stored in the AR does not represent the critical condition for a flare to occur, but it may be related to the size of the event once a flare does occur. Figures \ref{fig:f7} (e) and (f) show that the free energy for M6.6 and X2.2 flares, respectively, were also at different levels and that the free energy for the latter was almost twice that of the former. Thus, significant energy accumulated in the core of the AR from day to day. When M6.6 flare happened, the free energy decreased ($\approx$8 $\%$), showing that some of the energy had been released by the flare. A similar thing happened in the X2.2 flare: some of the energy was released and the free energy decreased. This energy evolution is consistent with previous results by \citet{sun12,asc14} and \citet{mal14}, although we obtained a smaller energy level than they found because we calculated the energy only for the core region. However, the general trend of the free energy was still increasing even after the flare happened, showing that using just the free energy profile to predict the occurrence of a flare is somewhat problematic. This is because we cannot differentiate the critical state required to trigger a flare during the continuous accumulation of energy in an AR, although the stored energy and helicity are related to the size and class of a flare once it occurs. Our result confirmed that non-potential energy does not directly provides necessary criterion for an eruption as it has been known in previous studies, e.g. in \citet{sch05} and \citet{jin09}. This is also consistent with previous simulation study by \citet{par17}. In this sense, analyzing the $\kappa_{T_c}^\ast$ profile can be more helpful for monitoring the condition necessary to trigger a flare and for evaluating how likely it is that an AR may erupt. Information about the free energy is useful for estimating the maximum size (class) of a flare that can be produced in an AR.

\section{SUMMARY AND CONCLUSION}
We have analyzed the evolution of magnetic twist, $\kappa_{T_c}^\ast$, and free energy for AR 11158. During the three days of its flare-active period, we found that the magnetic twist and $\kappa_{T_c}^\ast$ increased before M6.6 and X2.2 flares, and then decreased just after the flares. High-twist magnetic field accumulated in the core of the AR near the PIL before both the flares. We also found that the locations and shapes of the high-twist regions corresponded to the flare ribbons for both the flares, which shows that reconnection during the flares mostly involved high-twist fields in the AR.  We found that more high-twist field remained after X2.2 flare in the AR than after M6.6 flare. Accordingly, we suggest that the magnetic field in the AR relaxed considerably after M6.6 flare, but it retained its strong shear after X2.2 flare.

We have examined the values of $\kappa_{T_c}^\ast$ as a proxy for the parameter $\kappa$ in order to explore the possibility of using this parameter to analyze the condition required for solar eruptions. We found that the value of $\kappa_{T_c}^\ast$ appeared to reach a certain level before a flare happened. This value of $\kappa_{T_c}^\ast$ decreased significantly after both the flares, and to almost comparable levels. This shows that $\kappa_{T_c}^\ast$ provides a consistent behavior, so that it may be used to analyze the condition necessary for the occurrence of a solar flare. We suggest that the $\kappa_{T_c}^\ast$ value is related to the stability condition represented by the $\kappa$ parameter introduced by \citet{isi17}. We showed that the ratio of the high-twist flux compared to the total flux in the AR can be important to assess the possibility of a solar flare occurrence. It is noted that recently \citet{par17} found the importance of the ratio of the current helicity to the total helicity as a possible flare trigger from their parametric simulations. They speculated that this ratio may be related to the instability criterion of the torus instability. Although we investigated high-twist flux instead of helicity, since the helicity and the twist are related, some characteristics between their parameter and $\kappa$ may also be related. We suggest that during the eruption process, a double-arc structure can develop into a simple torus-like structure in the later phase. When it happens, it is possible that the torus instability will work and the TI threshold will determine the eruption.  

\begin{figure}
\epsscale{1.0}
\plotone{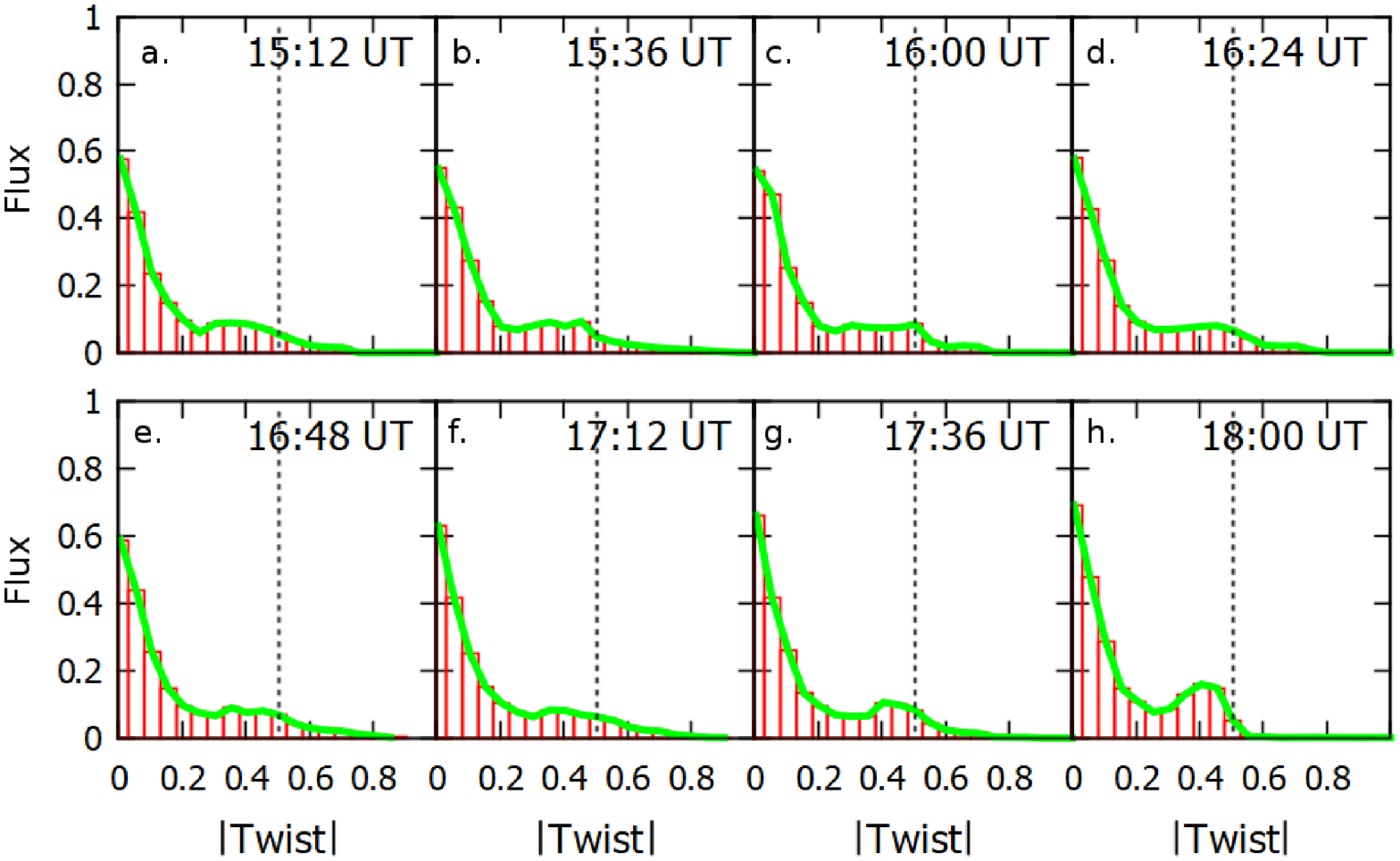}
\caption{Evolution of the twist distribution of AR 11158 before and after the M6.6 flare on 2011 February 13. The M6.6 flare onset time is 17:28 UT. The flux plotted on the y-axis is normalized by $7.72 \times 10^{21}$ Mx. The dashed vertical line in each panel marks the value of twist equal to 0.5 turn. 
\label{fig:f2}}
\end{figure}

\begin{figure}
\epsscale{1.1}
\plotone{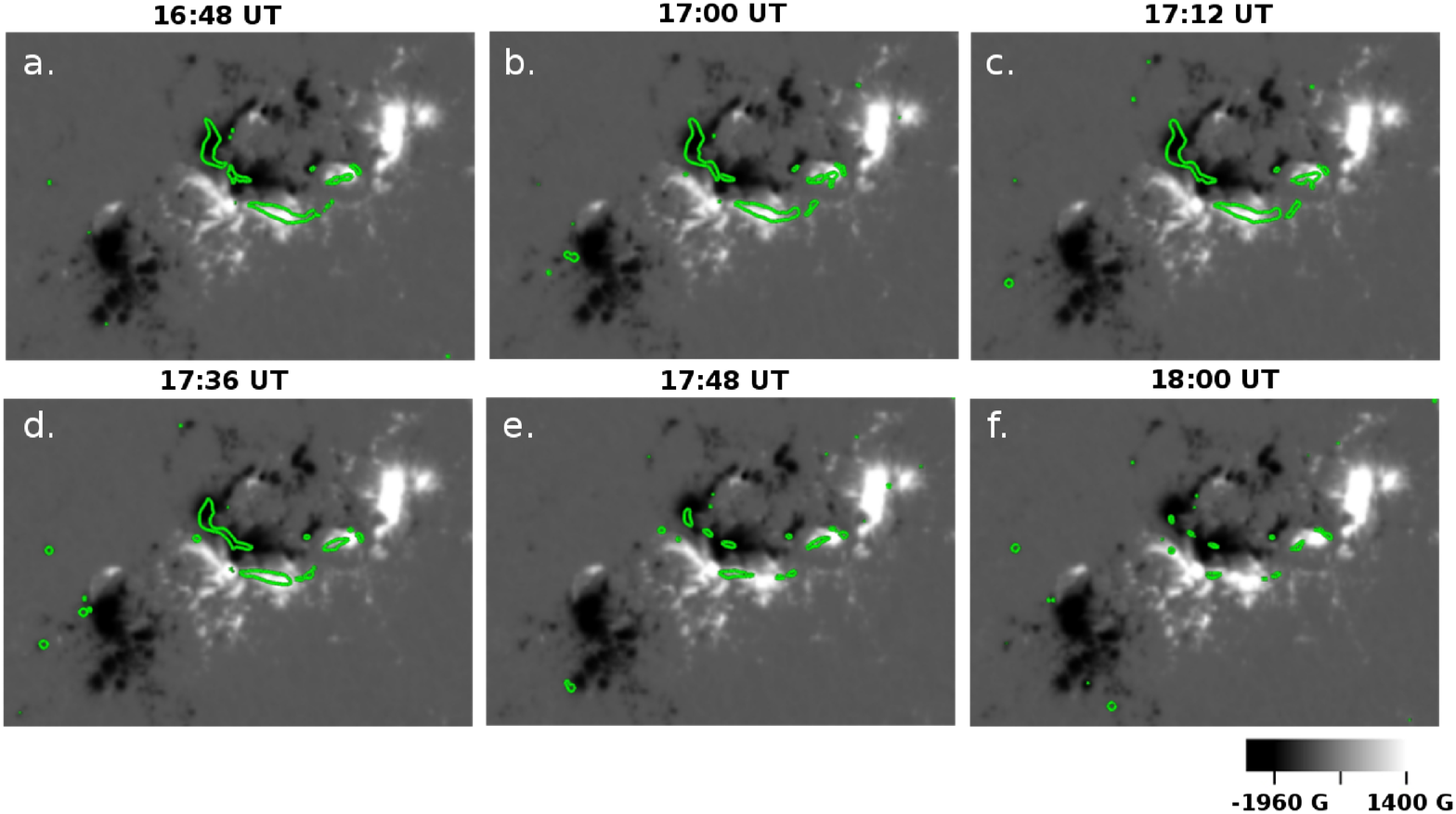}
\caption{Spatial evolution of the high-twist field in AR 1158 before (a, b, c) and after (d, e, f) M6.6 flare on 2011 February 13. The green contours correspond to areas with absolute magnetic twist equal to 0.5. The top (bottom) panels show the evolution of the high-twist distribution before (after) the M6.6 flare. 
\label{fig:f3}}
\end{figure}

\begin{figure}
\epsscale{1.0}
\plotone{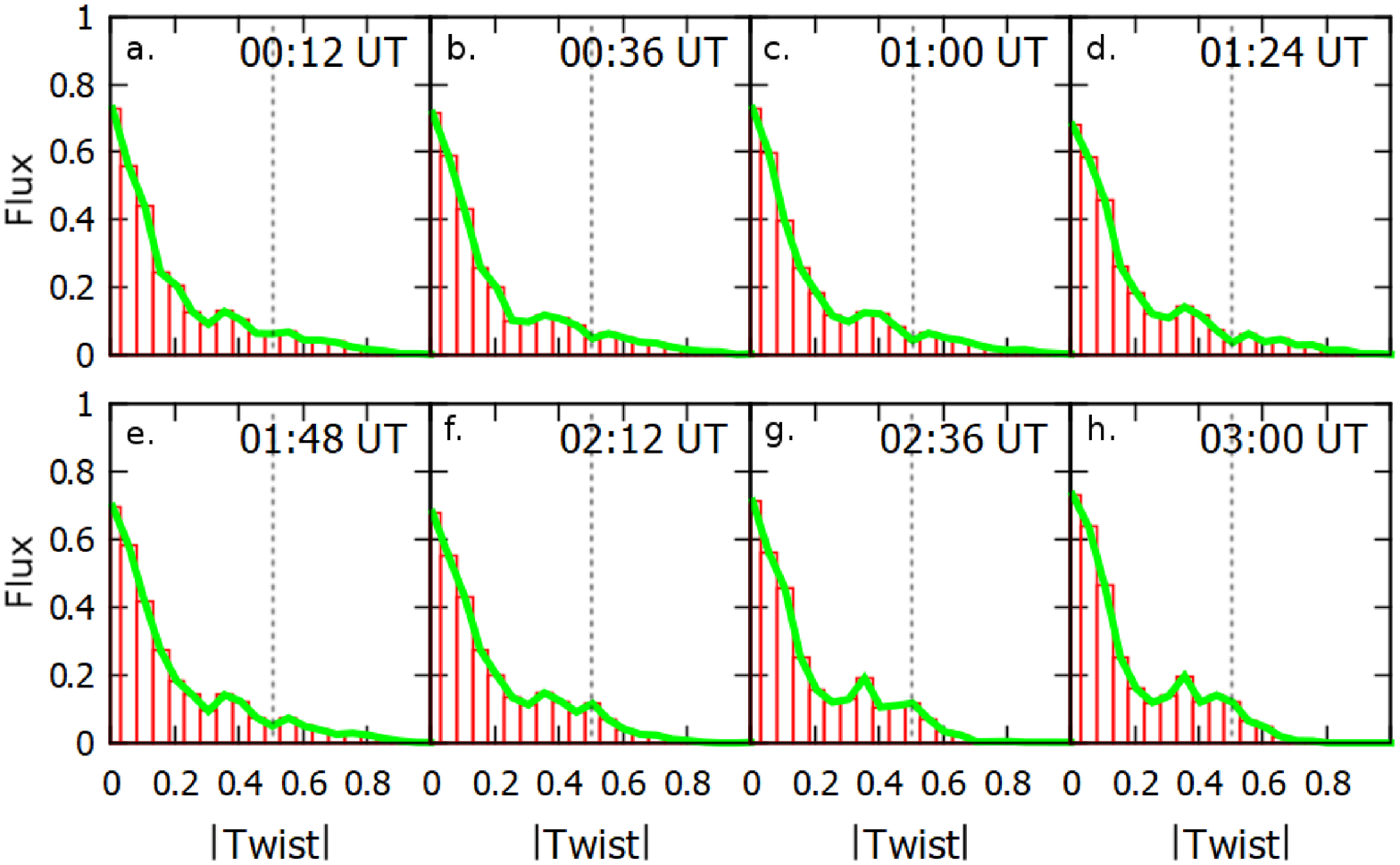}
\caption{Evolution of the twist distribution of AR 11158 before and after the X2.2 flare on 2011 February 15. The X2.2 flare onset time is 01:44 UT. The flux plotted on the y-axis is normalized by $7.72 \times 10^{21}$ Mx. The dashed vertical line in each panel marks the value of twist equal to 0.5 turn.
\label{fig:f4}}
\end{figure}

\begin{figure}
\epsscale{1.1}
\plotone{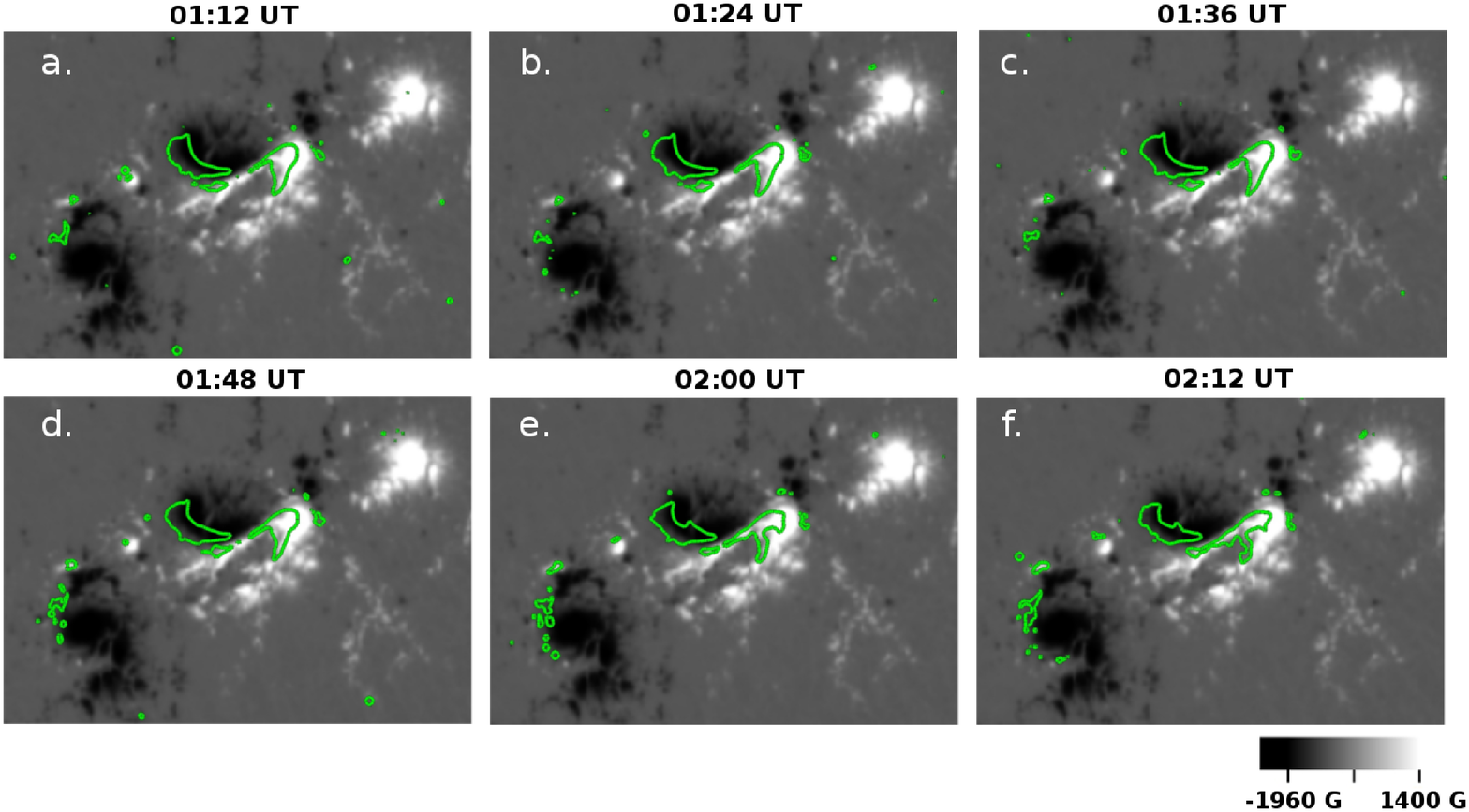}
\caption{Spatial evolution of high-twist field in AR 1158 before (a, b, c) and after (d, e, f) the X2.2 flare on 2011 February 15. The green contours correspond to areas with absolute magnetic twist equal to 0.5. The top (bottom) panels show the evolution of high-twist distribution before (after) the X2.2 flare. 
\label{fig:f5}}
\end{figure}

\begin{figure}
\epsscale{1.0}
\plotone{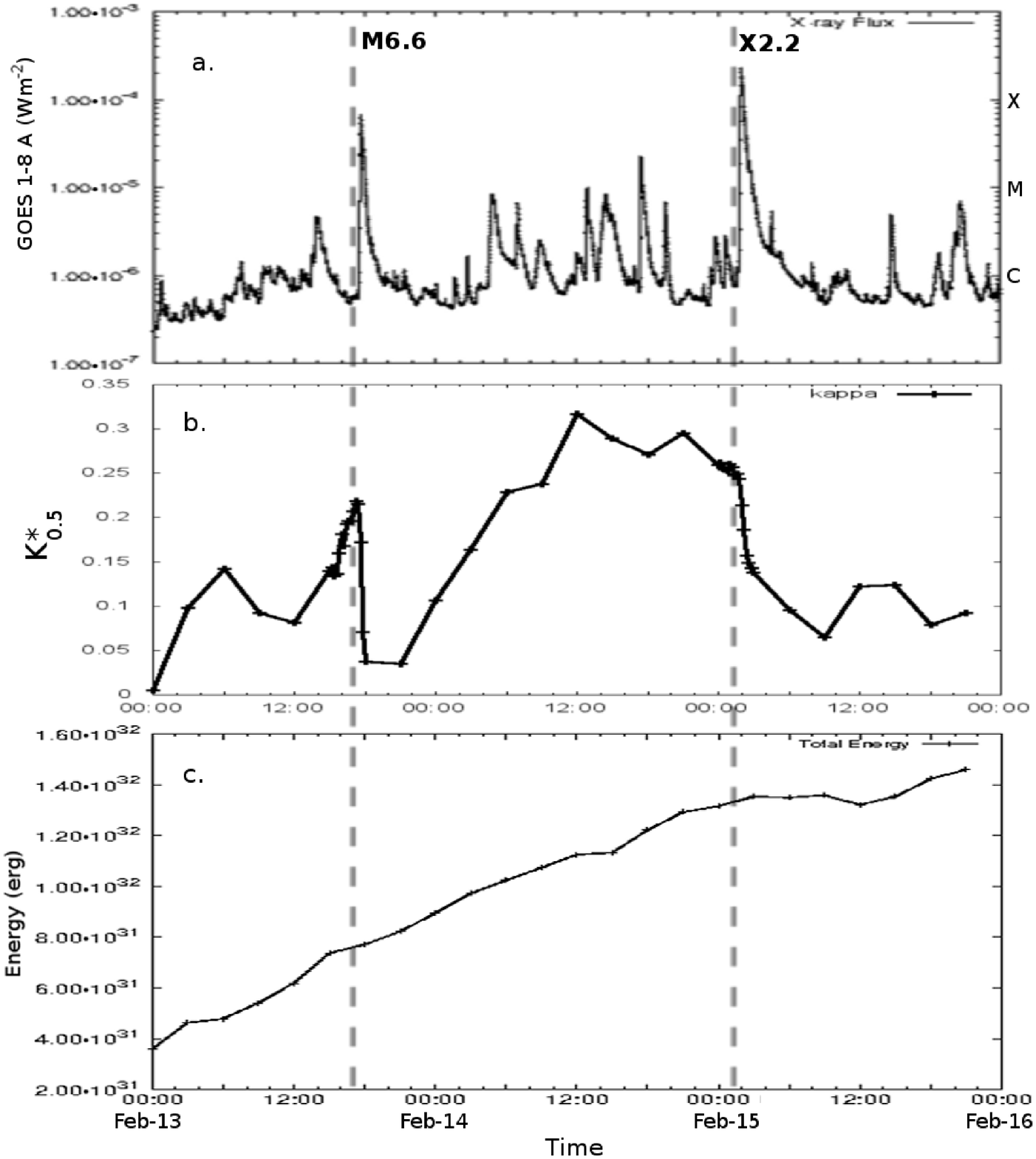}
\caption{Evolution of the GOES X-ray flux (a), $\kappa_{0.5}^\ast$ (b), and the total magnetic energy of the core of the AR (c) during the flare-active period from 2011, February 13 to February 15. The dashed vertical lines mark the onset times of the flares.
\label{fig:f6}}
\end{figure}

\begin{figure}
\epsscale{0.8}
\plotone{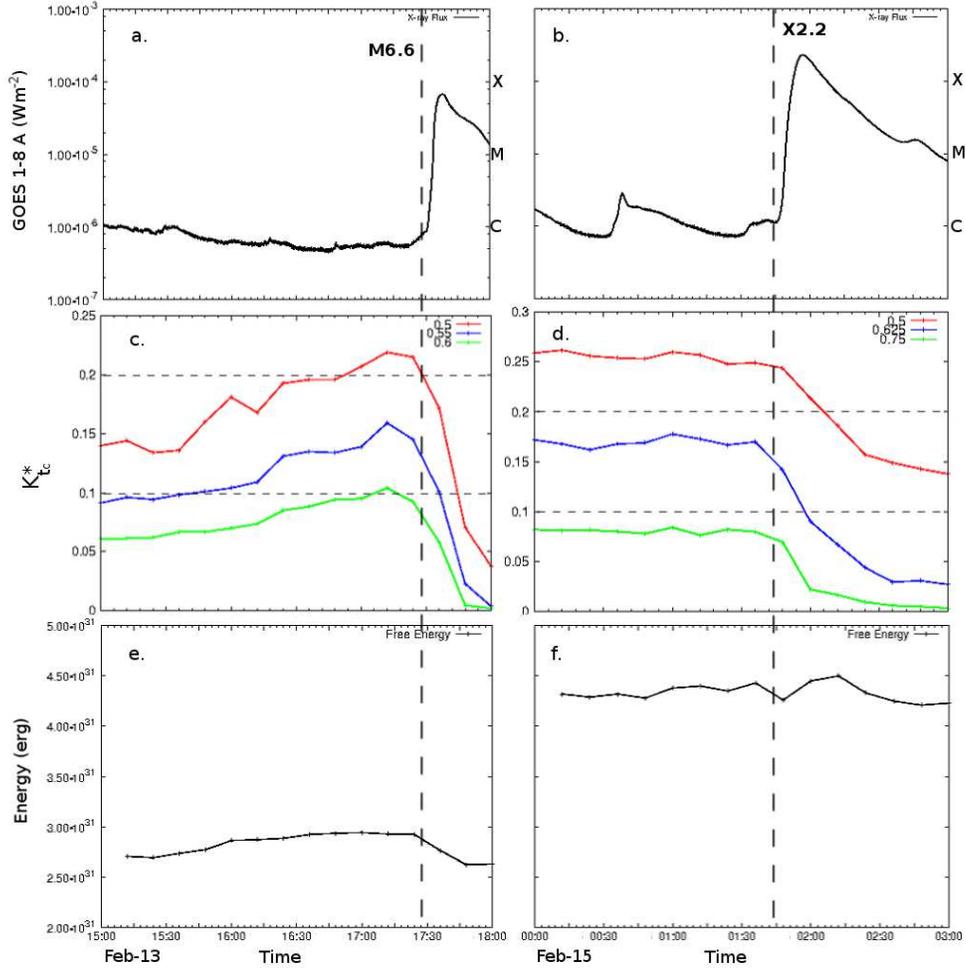}
\caption{Panels (a), (c), and (e) show the evolution of the GOES X-ray flux, $\kappa_{T_c}^\ast$, and the free energy of the core of the AR for the M6.6 flare, respectively. Panels (b), (d), and (f) show the corresponding quantities for the X2.2 flare, respectively. The legends in panels (c) and (d) show three values of $T_c$ as colored lines. The dashed vertical lines mark the onset times of the flares.
\label{fig:f7}}
\end{figure}

\begin{figure}
\epsscale{1.0}
\plotone{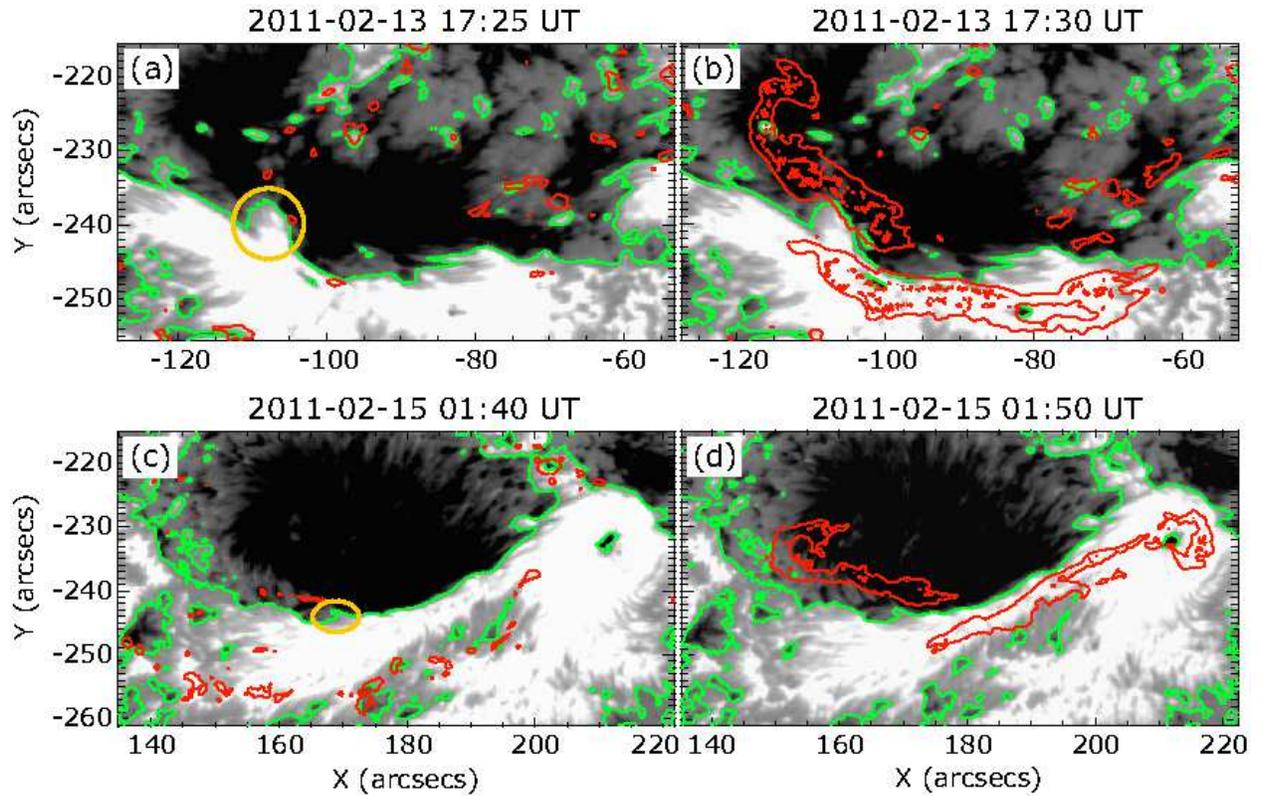}
\caption{Evolution of M6.6 flare (top) and X2.2 flare (bottom) plotted on the line-of-sight (LOS) component of the magnetic field obtained from Hinode/SP. The white and black colors correspond to positive and negative polarities, respectively. The green lines represent the PIL, and red contours show the strong Ca II H line emission. Yellow circles outline the magnetic ``peninsula" structures that trigger the flares.
\label{fig:f8}}
\end{figure}

Moreover, we found that $\kappa_{T_c}^\ast$ is very sensitive to the threshold ($T_c$) chosen to define the high-twist flux. Our analysis shows that the $\kappa_{T_c}^\ast$ value is more consistent if the threshold $T_c$ is chosen as the maximum twist of the flux remaining after a flare. However, we found that the $T_m$ value is not the same for different flares. Figures \ref{fig:f2} and \ref{fig:f4} show that the X2.2 flare had a larger value of $T_m$ than the M6.6 flare. It is still an open issue how to determine the threshold ($T_m$) before the onset of a flare in order to use $\kappa_{T_m}^\ast$ to predict a flare. One possibility is that $T_m$ may be related to the flare triggering process in which some disturbance of the magnetic field may play a role. It is likely that the value of $T_m$ should be higher (lower) if a smaller (larger) disturbance causes the triggering reconnection of less (more) flux. Further analysis of the evolution of $\kappa_{T_m}^\ast$ for many more flares and for non-flaring ARs is required to determine the proper value of $T_m$. However, the result of our study suggests that DAI analysis can be applied to a real AR in the Sun as long as we can obtain reliable values for the parameters required to calculate the $\kappa$. 

We also found that the magnetic energy decreased slightly before the flares, although it increased almost continuously during three days flare-active period. The free energy level in the AR was different for the two flares we studied. The AR had a higher free energy level before the X2.2 flare than before the M6.6 flare. However, we found that merely using the free energy in an AR is not sufficient to predict the onset of a flare, because it does not contain information about the stability of the magnetic system in the AR. The free energy level can indicate how large class of a flare can be produced by an AR, while $\kappa_{T_m}^\ast$ may be useful as a parameter to indicate an impending eruption in an AR. We believe that further studies based on the DAI analysis and its proxy parameter $\kappa_{T_m}^\ast$ can be helpful in achieving better understanding of the conditions necessary for a flare.

\acknowledgments

We thank all the team of SDO/HMI for producing vector magnetic field data products. We acknowledge Dr.\ K.D.\ Leka for fruitful discussion and comments that improved this paper. The authors are grateful to the anonymous referee for the constructive comments which lead to the improvement of the manuscript. Visualization of the NLFFF is produced by VAPOR \citep{cly07,cly05}. This work was supported by MEXT/JSPS KAKENHI Grant Numbers JP15H05814 and by MEXT as ``Exploratory Challenge on Post-K Computer" (Environmental Variations of Planets in the Solar System). JM acknowledges Indonesia Endowment Fund for Education (LPDP) for supporting his study and stay in Nagoya University. YB was supported by MEXT/KAKENHI Grant Number JP16H07478.

\clearpage

\end{document}